\documentstyle[preprint,aps,tighten,eqsecnum]{revtex}

\def\.{\!\cdot\!}
\def\[{\left[}
\def\]{\right]}
\def\({\left(}
\def\){\right)}
\def\bk#1{\langle#1\rangle}
\def\be{\begin{eqnarray}}
\def\ee{\end{eqnarray}}
\def\bm{\boldmath}
\def\ubm{\unboldmath}
\def\eq#1{(\ref{#1})}
\def\h{{1\over 2}}
\def\labels#1{\label{#1}}
\def\nn{\nonumber}

\def\a{\alpha}
\def\d{\delta}
\def\e{\epsilon}
\def\f{\phi}

\def\l{\lambda}
\def\p{\partial}
\def\r2{\sqrt{2}}
\def\rr2{{1\over\sqrt{2}}}

\def\t{\tau}
\def\v{\vec }

\def\Tr{{\rm Tr}}
\def\tr{{\rm tr}}
\def\un#1{{\bf #1}}
\def\uns{\un s}
\def\unt{\un t}
\def\unx{\un x}
\def\uny{\un y}
\def\unz{\un z}
\def\unk{\un k}
\def\vd{v^\dagger}

\def\ov{\overline}

\def\O{{\cal O}}
\def\D{{\cal D}}

\def\U{{\cal U}}
\def\A{{\cal A}}
\def\J{{\cal J}}

\begin{document}


\title{Saturation and Wilson Line Distribution}

\author{C.S. Lam$^a$, Gregory Mahlon$^b$,
and Wei Zhu$^c$\\}
\address
{\vspace*{.3cm}
$^a$Department of Physics, McGill University, \\
3600 University St., Montr\'eal, QC  Canada H3A 2T8\\
email: Lam@physics.mcgill.ca\\ \vspace*{.2cm} 
$^b$Department of Physics, Penn State Mont Alto,\\
1 Campus Drive, Mont Alto, PA 17237, U.S.A.\\
email: gdm10@psu.edu\\ \vspace*{.2cm}
$^c$Department of Physics, East China Normal University,\\
Shanghai 200062, China\\
email: Zhuweia@public8.sta.net.cn\\}
\maketitle
\begin{abstract}
We introduce a Wilson line distribution function $\ov W_\t(v)$ to
study gluon saturation at small Feynman $x_{_F}$, or large  $\t=\ln(1/x_{_F})$.
This new distribution can be obtained from the distribution
$W_\t(\a)$ of the Color Glass Condensate model and
the JIMWLK renormalization group equation. 
$\ov W_\t(v)$ is physically more relevant, and mathematically
simpler to deal with
because of unitarity of the Wilson line $v$.
A JIMWLK
equation is derived for $\ov W_\t(v)$; its properties are studied.
These properties
 are used to complete Mueller's derivation of the JIMWLK
equation, though for $\ov W_\t(v)$ and not $W_\t(\a)$. 
They are used to derive a generalized Balitsky-Kovchegov equation for
higher multipole amplitudes.
They are also
used to compute the unintegrated gluon distribution at $x_{_F}=0$,
yielding  a completely flat spectrum in transverse momentum
 squared $\unk^2$, with a known height. 
This is similar but not identical to the mean field result at
small $\unk^2$.
\end{abstract}

\section{Introduction}
Soft gluons 
are produced by radiation from more energetic partons.
Since the number of sources increases
at small Feynman $x_{_F}$, the soft gluon density $x_{_F}G$
per unit rapidity interval increases with $\t=\ln(1/x_{_F})$.
In fact, both the DGLAP equation \cite{DGLAP} and the BFKL 
equation \cite{BFKL} predict a growth so fast that the unitarity bound $\t^2$
is violated. To restore unitarity, a new mechanism is required to slow 
down the growth
towards $x_{_F}=0$ \cite{LG}.
The momentum $Q_s$ at which this mechanism sets in is known as the 
{\it saturation momentum}. 

The phenomenological implications for the presence of 
a saturation momentum have been
discussed in many papers \cite{EXPT}, but it is not yet
clear whether saturation has been observed experimentally.
On the theoretical side,  the important 
thing to note for our present discussion is that
soft gluons can be treated as a classical color
potential $\a(\v x)$, because of its large density at small $x_{_F}$.
In this background, energetic partons interact with  
soft gluons through their Wilson lines. It is this 
interaction that is responsible for the
saturation process.

With the replacement of soft gluons by a classical Yang-Mills
potential, the growth of soft gluon density is determined by 
the $\t$ dependence of the distribution function
$W_\t(\a)$. This dependence is given by a renormalization
group equation known as the JIMWLK equation \cite{JI26}.  

In this paper we introduce and study a distribution $\ov W_\t(v)$
for the Wilson line $v$. We will show that this distribution can be obtained
from $W_\t(\a)$, and that
it still satisfies the JIMWLK equation. Compared to
$W_\t(\a)$, it has the advantage of being more directly physically relevant,
because many physical processes can be described in terms of Wilson lines
or dipole amplitudes. Moreover,
the Wilson lines $v$ are unitary matrices, living on the color
group manifold which is compact. This compactness brings with it a number
of mathematical advantages not shared by $W_\t(\a)$, whose argument
$\a$ lives in a non-compact linear space. 

$\ov W_\t(v)$ can be used to derive a number
of properties not easily obtainable directly from $W_\t(\a)$.
We will use these properties to complete Mueller's
proof of the JIMWLK equation \cite{AM5}, though the proof is valid for
$\ov W_\t(v)$ and not  for
$W_\t(\a)$. They will be used to derive a generalized Balitsky-Kovchegov
(BK) equation for multipole amplitudes. We find that once the non-linear
BK equation for the dipole amplitude is solved, all subsequent multipole
amplitudes can be obtained by solving only linear equations. This means
saturation of the dipole amplitude automatically leads to saturation of
higher multipole amplitudes.

$\ov W_\t(v)$ will also be used to compute the asymptotic behavior of 
unintegrated gluon distribution at $x_{_F}=0$. 
We get a flat distribution
in the gluon transverse momentum $\unk^2$, with a height given by 
eq.~\eq{gluespec}. This is to be contrasted with the mean field
result \eq{meanfield} which yields a logarithmic dependence on
$\unk^2$ with an undetermined normalization.
Pre-asymptotic corrections will also be breifly discussed.

In the next section we start with a short review of saturation, the 
JIMWLK equation, the related BK
equation, the 
BFKL equation, as well as some of their solutions. In Sec.~3, 
the Wilson line distribution $\ov W_\t(v)$ is introduced,
together with some mathematical preliminaries. 
 The JIMWLK equation for $\ov W_\t(v)$
is derived, and its properties studied. In Sec.~4, the missing steps of
Mueller's derivation of the JIMWLK equation are supplied. In Sec.~5, 
the infrared divergence encountered in the JIMWLK equation
is discussed. It is shown that certain multipole functions which we call
multipole traces are free of these divergences. 
A generalized BK equation is derived to describe the evolution of
the multipole amplitudes.
In Sec.~6, we discuss
the property and implications  of the asymptotic solution of $\ov W_\t(v)$
at $x_{_F}=0$. In particular, the unintegrated momentum spectrum of the 
gluon density is derived. Correction to the asymptotic limit when
$x_{_F}\not=0$ is briefly discussed. Certain mathematical details  
can be found in Appendices A and B at the end.

After this paper was submitted for publication, we were informed by
the Referee of an interesting recent paper \cite{BIW} in which
$\ov W_\t(v)$ was also introduced and its JIMWLK equation derived.
It went on to give a random-walk interpretation of the JIMWLK
equations, with $\t$ playing the role of time, $v$ the position
in the group manifold, and $\a$ the velocity.

\section{A Brief Review}
The number of gluons emitted by 
a valence quark, per unit rapidity $\tau=\ln(1/x_{_F})$ and  per unit
transverse momentum squared $\un k^2$, is 
given in perturbation theory by $\a_sC_F/\pi\un k^2$,
where
$\a_s=g^2/4\pi$ is the QCD fine structure
constant. $C_F$ is the Casimir number in the fundamental 
representation, being
$(N_c^2-1)/2N_c$ for the gauge group $SU(N_c)$, and $\h N_c$ for
$U(N_c)$. A nucleus with atomic number $A$ has $AN_c$ valence
quarks, so its unintegrated gluon distribution is 
$dN/d\t d\un k^2\equiv d(x_{_F}G)/d\un k^2=\a_sC_F(AN_c)
/\pi\un k^2$.

When integrated,
this formula encounters an infrared divergence at small $\un k^2$,
brought about by
the long range gluon field of the 
unshielded valence quarks.
However, quarks are confined inside color-singlet
nucleons, so such a long
range force is absent
beyond the nucleon radius $a$. Thus
$x_{_F}G(x_{_F},Q^2)=(\a_sC_F(AN_c)/\pi)\ln(Q^2a^2)$.

In the central rapidity region where Feynman $x_{_F}$ is small,
the gluon density is much larger than the amount given
by the perturbation formula. This is so because soft gluons can 
be radiated also from energetic gluons and sea quarks, not just the
valence quarks considered so far. 
According to the DGLAP \cite{DGLAP} equation, the soft gluon
density grows like $\exp(\kappa\sqrt{\t})$ for some positive constant
$\kappa$, and, according to the BFKL equation \cite{BFKL}, it grows like
 $\exp(4\a_s\ln(2)N_c\t/\pi)$. Both exceed the unitarity limit $\t^2$,
so a new mechanism must kick in to dampen the growth and
restore unitarity at small $x_{_F}$ \cite{LG}. This effect is known as {\it saturation}. 

Saturation is thought to arise from a non-linear mechanism which occurs 
when gluons are sufficiently dense to interact among themselves 
\cite{AM4,EI1}. The number of gluons per unit rapidity interval is
$x_{_F}G$. In a nucleus of radius $R_A$,
the transverse area per gluon is therefore $\pi R_A^2/x_{_F}G(x_{_F},Q^2)$.
The average color-charge squared of a gluon
is $N_c/(N_c^2-1)$, their interaction strength is $\sim\a_s/\pi$,
and their natural size is $\sim 1/Q$. Hence the cross section for
two gluons to interact can be estimated to be $(\a_s/\pi)(N_c/(N_c^2-1))
(\pi/Q^2)$. If the cross section is larger than the transverse
area per gluon, then interaction will take place to
set off the non-linear mechanism. The onset therefore occurs at
a momentum scale $Q_s$ such that
\be
\pi R_A^2/x_{_F}G(x_{_F},Q_s^2)=c(\a_s/\pi)(N_c/(N_c^2-1))(\pi/Q_s^2),\ee
or equivalently,
\be
Q_s^2=c {\a_sN_c \over N_c^2-1}{x_{_F}G(x_{_F},Q_s^2)
\over\pi R_A^2}.\labels{qs}\ee 
A constant $c$ has been inserted to account for the qualitative nature
of this argument. Even when $Q_s^2$ is obtained from a detailed calculation,
the constant $c$ is still somewhat ambiguous  
because transition into saturation does not occur sharply. 
Thus one finds a number of $c$'s used in the literature.
For example, $c=1$ in \cite{EI5},
$c=\pi^2$ in \cite{EI1} when estimated from the mean field approximation
on the large $\unk^2$ side, and $c=16\pi^2c_1$ when estimated
on the small $\unk^2$ side, where $c_1$ is some unknown constant. 
And, $c=\pi$ in \cite{EI3}. 

$Q_s$ can also be defined through the unintegrated gluon density
$d(x_{_G}G)/d\un k^2$.
When we reduce $\unk^2$ from infinity, this density increases
until a point $\unk^2=Q_s^2$
when gluons become sufficiently dense to set off the non-linear mechanism.
From there on we enter a saturation region with a much slower growth.
However, this definition is also ambiguous unless the slowdown occurs
fairly sharply, which turns out to be the case at $x_{_F}=0$.
As we shall see in Sec.~6A, at $x_{_F}=0$,
the saturation region is large and the unintegrated spectrum 
$d(x_{_F}G)/d\unk^2$ 
in this region is absolutely flat in $\unk^2$. 
This then allows  $c$ to be 
determined unambiguously to be $c=8\pi^3$.

Using the BFKL solution for $x_{_F}G$ as a qualitative estimate, 
and assuming that $x_{_F}G$ is proportional to $A$, we see from \eq{qs} that
$Q_s^2$ grows with a power of $1/x_{_F}$ and $A^{1/3}$, 
making it large for large nucleus or small $x_{_F}$. \eq{qs}
also implies that the gluon number per unit transverse area
at saturation is $\sim Q_s^2(x)/\a_s$.

The large number of gluons present at saturation
allows them to be 
treated as a classical Yang-Mills (YM) potential $\a^a(\v x)$.
The superscript $a$ is the color index, and $\v x=(x^-,\unx)$ are the lightcone 
(LC) coordinates, defined for a hadron moving  
along the $+z$ direction to be
$x^\pm=(t\pm z)/\sqrt{2}$ and
$\unx=(x^1,x^2)$. It is also convenient to introduce the spacetime
rapidity variable $y=\ln(x^-P^+)$, where $P^+$ is the 
$+$ component of the hadron momentum,
and the gluon potential $\a_y^a(\unx)=x^-\a(x^-,\unx)$.

For a fast moving hadron (or nucleus), 
Lorentz contraction forces $\a(\v x)$ to be
concentrated around $x^-=0$, and time dilation makes it effectively
(LC) time ($x^+$) independent. The soft gluons are produced by
partons within the hadron, so one can assume $\a_y^a(\unx)=0$
for $y>\t$ \cite{EI1}.

Energetic partons, whether in the same hadron or not, interact with the soft gluons through the Wilson-line factors
\footnote{
We shall use upper
case letters to denote the adjoint representation and lower
case letters to denote the defining representation. In this notation,
the generators in the defining representation will be denoted
by $t_a$, and they will be normalized to be $\tr(t_at_b)=\h\d_{ab}$.
The generators in the adjoint representation are denoted by
$T_a$. They are related to the totally antisymmetric structure
constants by $(T_a)_{bc}=if_{bac}$. Hence $(T_a)_{bc}$ is imaginary
and totally antisymmetric in the three indices. Similarly,
the quark and anti-quark Wilson lines will be denoted by the lower-case
letters $\vd$ and $v$, and the gluon Wilson line will be denoted by
the upper-case letters $V^\dagger$ and $V$. A slight drawback of this
convention is that we are forced to denote the dipole amplitude
\eq{dipole} by a lower-case letter $s_\t$, whereas the usual notation
for it is $S_\t$.} 
\be
\vd(\un x)&=&P\exp\(+ig\int_{-\infty}^\infty dx^-\a^a(\v x)t_a\)
=P\exp\(+ig\int_{-\infty}^\t dy\ \a^a_y(\unx)t_a\),\nn\\
v(\un x)&=&
\widetilde P\exp\(-ig\int_{-\infty}^\infty dx^-\a^a(\v x)t_a\)
=\widetilde P\exp\(-ig\int_{-\infty}^\t dy\ \a^a_y(\unx)t_a\),\nn\\
V^\dagger(\un x)&=&P\exp\(+ig\int_{-\infty}^\infty dx^-\a^a(\v x)T_a\)
=P\exp\(+ig\int_{-\infty}^\t dy\ \a^a_y(\unx)T_a\),
\nn\\
V(\un x)&=&
\widetilde P\exp\(-ig\int_{-\infty}^\infty dx^-\a^a(\v x)T_a\)
=\widetilde P\exp\(-ig\int_{-\infty}^\t dy\ \a^a_y(\unx)T_a\).
\labels{wilson}\ee
where $P$ and $\widetilde P$ indicate respectively path-ordering and
anti path-ordering. The first two expressions describe the propagation of
quarks and anti-quarks, respectively,
through the dense background of the soft gluons,
 and the last two expressions describe
the propagation of gluons. These Wilson lines play a central role in the
rest of the paper.

In this representation of soft gluons by a classical background field,
gluon distribution is determined by the distribution
$W_\t(\a)$ of the YM potential. 
$W_\t$ depends on $\t$ because 
the number of sources available to emit soft gluons 
increases at small $x_{_F}$.

The resulting change of the distribution functional
$W_\t(\a)$ can be shown to satisfy
the JIMWLK renormalization group equation \cite{JI26}
\be
{\p W_\t(\a)\over\p\t}=-HW_\t(\a),\labels{jimwlk1}\ee
where
\be
H&=&{1\over\pi}\int d^2\unz d^2\unx d^2\uny
K(\unx\un y|\un z)\O(\unx\un y|\un z),\nn\\
\O(\unx\un y|\un z)&=&
{\d\over\d\a^a_\t(\unx)}\[V^\dagger(\unx)-V^\dagger(\un z)
\]_{ac}\[V(\un y)-V(\un z)\]_{cb}{\d\over\d\a^b_\t(\un y)},\nn\\
&&\labels{jimwlk2}\ee
and
\be
K(\unx\un y|\un z)&=&{1\over 4\pi^3}{(\un x-\un z)
\.(\un y-\un z)\over(\un x-\un z)^2(\un y-\un z)^2}.\labels{jimwlk3}\ee

A consequence of \eq{jimwlk1} and \eq{jimwlk2} is that
the normalization $\int \D[\a]W_\t(\a)$ is independent
of $\t$. We will normalize it to be 1, so that
the average of any functional of $\a$
is  given simply by $\bk{F}_{_\t}=\int \D[\a]F(\a)W_\t(\a)$.

The functional derivatives $\d/\d\a_\t^a$ of $V^\dagger$ and $V$
in \eq{jimwlk2},
and similarly of $\vd$ and $v$ that we will encounter later, 
are
\be
{\d V^\dagger(\un z)\over\d\a_\t^a(\un x)}&=&igT_aV^\dagger(\un z)
 \d(\unx-\un z),\nn\\
{\d v^\dagger(\un z)\over\d\a^a_\t(\un x)}&=&igt_av^\dagger(\un z)
\d(\unx-\un z),\nn\\
{\d V(\un z)\over\d\a_\t^a(\un x)}&=&-igV(\un z)T_a\d(\unx-\un z),\nn\\
{\d v(\un z)\over\d\a_\t^a(\un x)}&=&-igV(\un z)t_a\d(\unx-\un z).
\labels{gluewil4}\ee

For calculational simplicity it is useful to note that
$V^\dagger(x)-V^\dagger(z)$ in \eq{jimwlk2} may be put
in front of the operator $\d/\d\a_\t^a(\unx)$. This follows from
\eq{gluewil4} and the observation that $(T_a)_{ac}=if_{aac}=0$. 

A particularly important physical quantity to study is the dipole amplitude
\cite{AM}
\be
s_\t(\unx,\un y)={1\over N_c}\bk{\tr\[\vd(\un x)
v(\un y)\]}_{_\t}.\labels{dipole}\ee
At coincident points
\be
s_\t(\unx,\unx)=1\labels{normdip}\ee
 because of unitarity of $v$. 
It can be shown form \eq{jimwlk1} that the dipole amplitude satisfies
the Balitsky equation \cite{IB}
\be
{\p s_\t(\unx,\un y)\over\p\t}&=&-{\a_sN_c\over 2\pi^2}
\int d^2\unz {(\unx-\un y)^2\over (\unx-\un z)^2(\un y-\un z)^2}\.\nn\\
&\.&
\Bigg\{s_\t(\unx, \un y)-{1\over N_c^2}\bk{\tr\[\vd(\unx)v(\un z)\]
\tr\[\vd(\un z)v(\un y)\]}_{_\t}\Bigg\}.\nn\\
\labels{balitsky}\ee
Note that the infrared divergence occuring at large $\un z$ in 
\eq{jimwlk2} is absent in the Balitsky equation.

For large $N_c$, the last term factorizes and we arrive at the
Kovchegov equation \cite{YK}
\be
{\p s_\t(\unx,\un y)\over\p\t}&=&-{\a_sN_c\over 2\pi^2}
\int d^2\unz{(\unx-\un y)^2\over (\unx-\un z)^2(\un y-\un z)^2}\.\nn\\
&\.&
\Bigg\{s_\t(\unx, \un y)-s_\t(\unx,\un z)s_\t(\un z,\un y)\Bigg\}.
\labels{kovchegov}\ee

Since $s_\t(\unx,\unx)=1$, the quantity $t_\t(\unx, \un y)\equiv
1-s_\t(\unx, \un y)$ is expected
to be small when $\unx\sim\un y$. If we deal with soft gluons outside
of the saturation region, when $\a(\v x)$ is small, the Wilson lines
$v$ and $\vd$ are close to 1 anyway, so we expect to be able to drop
the quadratic term
$t_\t(\unx,\unz)t_\t(\unz,\uny)$. In this way we get the 
dipole form of the BFKL equation
\cite{BFKL}
\be
{\p t_\t(\unx,\un y)\over\p\t}&=&-{\a_sN_c\over 2\pi^2}
\int d^2\unz{(\unx-\un y)^2\over (\unx-\un z)^2(\un y-\un z)^2}\.\nn\\
&\.&
\Bigg\{t_\t(\unx, \un y)-t_\t(\unx,\un z)-t_\t(\un z,\un y)\Bigg\}.
\labels{bfkl}\ee

For $|\unx-\uny|^2\gg 1/Q_s^2$ and inside the saturation region,  
the strong classical YM potential $\a$
causes large and independent oscillations to both
Wilson lines $\vd$ and $v$. Consequently the dipole amplitude
$s_\t(\un x-\un y)$ is expected to be small, thus enabling the non-linear
 term in \eq{kovchegov} to be dropped. 
The solution of the resulting linear equation can be
shown to be \cite{AM4}
\be
s_\t(\un x-\un y)=\exp\[-{\a_sN_c\over\pi}\int_{\t_0}^\t
dy\ln(Q_s^2(y)\un x^2)\]s_{\t_0}(\un x-\un y).\labels{larges1}\ee
If $Q_s^2(y)=
\exp[c\a_sN_c(y-\t_0)/\pi]Q_s^2(\t_0)$ \cite{QSEST}, then
\be
s_\t(\un x-\un y)=\exp\[-{c\over 2}\({\a_sN_c\over\pi}\)^2(\t-\t_0)^2\]
s_{\t_0}(\unx-\un y),\labels{larges2}\ee
provided $\t_0$ is chosen so that $Q_s^2(\t_0)(\unx-\un y)^2=1$.
This condition implies $(\unx-\un y)^2=Q_s^{-2}(\t_0)\gg Q_s^{-2}(\t)$,
if $\t\gg\t_0$. The solution \eq{larges2} then confirms the expectation
that $s_\t$ is small in that region.

To solve any of these evolution equations we need an initial
condition at some $\t=\t_0$.
For a large nucleus, and a $\t_0$ where the source is dominated
by the valence quarks, the initial condition is provided by 
the McLerran-Venugopalan model \cite{MV}, in which
a Gaussian distribution is assumed for $W_{\t_0}(\a)$. 
For large $A$ and small
$\a_s(Q_s^2)$, the Gaussian distribution
 can be shown to be a good approximation \cite{LM3}.
Saturation is now provided by the valence quarks alone,
so $A$ has to be very large and the resulting $Q_s^2$ is relatively small.
The detail of confinement which affects the small $\unk^2$ region
then becomes relatively important \cite{LM12}.

With the Gaussian distribution, the perturbative gluon distribution
is modified by a gluon dipole factor \cite{EI1}
\be
S_{x^-}(\un r)={1\over N_c^2-1}\bk{\Tr U^\dagger(x^-,\un r)U(x^-,\un 0)},
\labels{gluedipole}\ee
where $U^\dagger$ is equal to $V^\dagger$ in \eq{wilson} with
$\t$ replaced by $x^-$, and $\un r$ is the conjugate
variable to the transverse momentum $\un k$. This factor
gives rise to the non-linear effect that is responsible
for saturation, with a saturation momentum 
$Q_s$ given by \eq{qs}.

For $\t>\t_0$, the $W_\t(\a)$ determined by \eq{jimwlk1} no longer has
a Gaussian distribution. Nevertheless, in a mean field approximation,
the approximate solution is still  Gaussian. The gluon distribution
for small $\un k^2$ is then given by \cite{EI1}
\be
{d(x_{_F}G)\over d^2\un k}\simeq c_2{N_c^2-1\over N_c}{\pi R_A^2\over\a_s}
\ln{Q_s^2(\t)\over\un k^2},\quad (\un k^2\ll Q_s^2(\t))\labels{meanfield}\ee
for some constant $c_2$.

\section{Wilson Line distribution}
We shall show later in this section that 
the distribution $W_\t(\a)$ of YM potential $\a$ leads to a distribution 
$\ov W_\tau(v)$ of the Wilson lines.

The notation might suggest that $\ov W_\t(v)$ describes only the
 distribution of anti-quark Wilson
line $v$, but actually it provides a distribution for the 
Wilson lines of other
partons as well. Since $v$ is unitary, $\vd=v^{-1}$,
the variable $\vd$ is  a 
rational function of $v_{ij}$, so $\ov W_\t(v)$ does provide the distribution
for quark Wilson lines $\vd$. 

From the group-theoretical relation
\be
t_aV^\dagger_{ac}=\vd t_cv=V_{cb}t_b,\labels{vv1}\ee
or equivalently, the relation
\be
V^\dagger_{ac}=2\tr(\vd t_cvt_a)=V_{ca},\labels{vv2}\ee
gluon Wilson lines can be expressed as a quark anti-quark
pair of Wilson lines, so we can also compute the distribution
of gluon Wilson lines from $\ov W_\t(v)$.

Physical observables are often expressible in terms of
Wilson lines, so it is clearly desirable
to know their distributions directly.
Moreover, the Wilson line $v$ is unitary, which allows
the theory of representation of the 
unitary group to be used for computations.
For example, an orthonormal complete set of polynomials,
given by the irreducible representations of the unitary group,  exists
on the group manifold. Therefore a harmonic analysis of
 $\ov W_\t(v)$ and other
physical quantities can be carried out to allow their integrals 
to be computed.
In contrast,
$W_\t(\a)$ is a function on a non-compact linear space of the
Lie algebra, and the only functional that can be integrated
in practice is the Gaussian. 

We shall show that $\ov W_\t(v)$ satisfies the same JIMWLK equation
\eq{jimwlk1}, but with $H$ replaced by
$\ov H$. The latter is obtained from $H$ simply by replacing the
$\a$ derivative by a differential operator in $v$. Therefore,
$\ov H$ is still hermitean and positive semi-definite. 

Since the group
manifold is compact, the spectrum and eigenfunctions of
$\ov H$ are more manageable. For example, $\ov W_\t(v)=1$ 
is a normalized
eigenfunction of $\ov H$ with zero eigenvalue, whereas $W_\t(\a)=1$
has a divergent integral in the $\a$ space. 

Before showing how $\ov W_\t(v)$ is obtained from $W_\t(\a)$,
let us first review some basic facts about integrations and 
orthonormal relations on a compact Lie group.

\subsection{Inner products on the color group}
For the sake of definiteness we shall assume the color group to be $U(n)$, 
though a similar analysis can be carried out for $SU(n)$. The 
$n^2$-dimensional group $U(n)$ 
will be parameterized by the $n^2$ matrix elements $v_{ij}
\ (1\le i,j\le n)$  in the defining representation. Unitarity
 equates $v^\dagger$ to $v^{-1}$, so $v^*_{ij}$ is to be 
regarded as a dependent variable, given as
 a rational function of $v_{kl}$ via this relation.

The invariant volume element on $U(n)$ will be denoted by $d_H[v]$. 
This Haar measure is a left and right 
invariant $n^2$-form, satisfying 
\be
d_H[v]=d_H[v_0v]=d_H[vv_0]\labels{haar}\ee
for any constant $v_0\in U(n)$. It is positive, and it
shall be normalized to 
$\int d_H[v]=1$.

The Haar measure $d_H[v]$ is proportional
but not equal to
the product measure $d[v]$, obtained by taking the exterior product of the
$n^2$ 1-forms $dv_{ij}$. They differ by a Jacobian $J(v)$, so
\be
d_H[v]=J(v)d[v].\labels{jac}\ee

To get an idea how this comes about, consider a change $dv$ in the vicinity
of a group element $v\in U(n)$. Then $v^{-1}dv$ constitutes a change
around the identity, so it can be parametrized
 in the form $-it_a(d\eta_a)$.
The volume element at the identity is proportional to the exterior
product of the $n^2$ $d\eta_a$'s, or equivalently, the exterior product
of $2i\tr(v^{-1}dv)$. If we want the volume element to be left and right
invariant, as in \eq{haar}, this expression should be taken as
the volume element at any point $v\in U(n)$. The presence of $v^{-1}$
in this expression is the origin of the Jacobian $J(v)$ in \eq{jac}.
When $n$ is odd, $d_H[v]$ has a very simple analytical form, given by
\eq{oddn} in Appendix A.

Let $f(v)$ and $g(v)$ be two functions on the group manifold. Their
inner product is defined to be
\be
\bk{f(v)|g(v)}=\int d_H[v]f^*(v)g(v).\labels{innergp}\ee
It will be shown later that the operator $\ov \D=v_{ij}(\p/\p v_{ij})$
is hermitean with respect to this inner product. Since
$\ov \D v_{kl}=v_{kl}$, 
the eigenfunctions
of $\ov \D$ are monomials of $v$, whose eigenvalues
are the degrees of the monomials. The inner product of two eigenfunctions
of a hermitean operator is zero if their eigenvalues are different. Hence
if $M_k(v)$ denotes a monomial
of degree $k$ and $M'_l(v)$ a monomial of degree $l\not=k$,
then 
\be
\bk{M_k(v)|M'_l(v)}=0\quad (k\not=l).\labels{ortho}\ee
This result can be stated in another way. An integral
$\int d_H[v]B(v,\vd)$ is non-zero only when the number of 
$v$'s in $B$ is equal to its number of $\vd$'s. We shall refer
to this later as the {\it matching rule}. It is one of the main
tools for our later calculations.

To compute inner products when $k=l$, we resort to the theory
of representation of the $U(n)$ group, which asserts
that if $D^\l_{CD}(v)$ is the
matrix element of an irreducible representation $\l$, then
\be
\int d_H[v]D^{*\l'}_{AB}(v)D^\l_{CD}(v)=
{1\over N(\l)}\d_{\l\l'}\d_{AC}\d_{BD},\labels{group}\ee
where $N[\l]$ is the dimension of the irreducible representation $\l$.

For example, the defining representation $v$ is irreducible
and has dimension $N_c$, hence
\be
\bk{v_{ij}|v_{kl}}={1\over N_c}\d_{ik}\d_{jl}.\labels{vv}\ee

The orthonormal relation \eq{group} can be used to compute inner products
of any two monomials in the following way. First, apply Young's idempotent
operators of $k$ boxes to decompose $M_k(v)$ into a linear combination
of irreducible representations $D^\l(v)$. Similarly $M'_l(v)$ is decomposed
into a linear combination of $D^{\l'}(v)$. Then
\eq{group} can be used to calculate $\bk{M_k(v)|M'_l(v)}$.
Clearly it can also be used to calculate integrals $\int d_H[v]B(v,\vd)$.

\subsection{Measures and functionals on $\A$ and $\U$}
To apply these properties to the physical problem on hand, 
we need to generalize them to the case when $v$ depends on the 
transverse position $\unx$. We shall denote the color
group containing $v(\unx)$ as $U(n)_{\un x}$, and 
${\cal U}\equiv\prod_{\un x}U(n)_{\un x}$.

The product over $\unx$ is to be interpreted in the following way.
Cover the transverse $\un x$-plane 
by a square lattice with a lattice constant $a$.
The product $\unx$ is to be taken over all lattice points within
the Lorentz-contracted
nucleus of transverse radius $R_A$. The same convention will be applied
to sums over $\unx$.

The measure on ${\U}$ is defined to be 
\be
\D_H[v]=\prod_\unx d_H[v(\un x)],\labels{bighaar}\ee
where $d_H[v(\un x)]$ is the Haar measure on $U(n)_{\un x}$.
Using \eq{jac} and denoting $\prod_\unx J(v(\unx))$ by $\J(v)$, we get 
\be
\D_H[v]=\J(v)\D[v],\nn\\
\D[v]\equiv \prod_\unx d[v(\unx)].\labels{dv}\ee

The measure $\D[\a]$ on the Lie algebra $\A$ of YM potentials 
can be defined in the following way. Divide the $y$ axis into intervals
of size $\e$. Since we are interested in $\a_y^a(\unx)$
only for $y\le \t$, the appropriate measure is
\be
\D[\a]&=&\D_\t[\a]\D_<[\a],\ee
where
\be
D_\t[\a]&\equiv&\prod_{a,\unx} d\a_\t^a(\unx),\nn\\
D_<[\a]&\equiv&c_3\prod_{y\le\t-\e}\prod_{a,\unx}
 d\a_y^a(\unx),\labels{da}\ee
with a normalization constant $c_3$ to be chosen later.
In the same vein, the Wilson line $v(\unx)$ of \eq{wilson}
can be factorized into 
\be
v(\unx)=v_<(\unx)v_\t(\unx),\labels{ufac}\ee
with
\be
v_<(\unx)&=&\widetilde P\exp\(-ig\int_{-\infty}^{\t-\e} dy\
\a_y^a(\unx)t_a\),\nn\\
v_\t(\unx)&=&\exp\(-ig\a_\t^a(\unx)t_a\e\).\labels{ufac2}\ee
A change of $\a_\t^a(\unx)$ of amount $d\a_\t^a(\unx)$ causes
a change of $v_\t(\unx)$ of amount $d v_\t(\unx)=-ig\e v_\t(\unx)
t_ad\a_\t^a(\unx)$,
and hence a change in $v(\unx)$ by the amount 
$d v(\unx)=-ig\e v(\unx)t_ad\a_\t^a(\unx)$. Or,
$v^{-1}d v(\unx)=-ig\e t_ad\a_\t^a$. In light
of the remark below \eq{jac}, we can now choose the constant $c_3$
in \eq{da} so that 
\be
\D_H[v]=\D_\t[\a].\labels{dvda}\ee

We are now ready to discuss how $\ov W_\t(v)$ can be obtained
from $W_\t(\a)$.

A functional of $\a_y(\unx)$ for $y\le\t$
 can be folded into a functional of $v(\unx)$ 
using the formula
\be
\ov F(v)&=&\int F(\a)\d\(v-u\){1\over \J(u)}\D[\a]\nn\\
&=&\int F(\a)\d(v-u)\D[u]\D_<[\a],\labels{map}\ee
where 
\be
u(\unx)=\widetilde P\exp\(-ig\int_{-\infty}^\t dy\ \a_y^a(\unx)t_a\),
\labels{u}\ee
 and 
\be
\d(v-u)\equiv \prod_\unx\prod_{i,j=1}^n\d[v(\unx)_{ij}-u(\unx)_{ij}],\ee
so that $\int f(u)\d(v-u)\D[u]=f(v)$ for any functional $f(u)$.
The second equality of \eq{map} comes from \eq{dv},
\eq{da}, and \eq{dvda}.

It follows from \eq{map} that
\be
\int \ov F(v)\D_H[v]&=&\int \D_H[v]\D[\a]\d(v-u)F[\a]/\J(u)\nn\\
&=&
\int F(\a)\D[\a],\labels{wint}\ee
so $\ov W_\t(v)$ is normalized if $W_\t(\a)$ is.

It also follows from \eq{map} by an integration by parts that the
transform of $\d F(\a)/\d\a_\t^a(\un x)$ is 
$\widehat \D_a(\unx) \ov F(v)$, where
\be
\widehat \D_a(\unx)\equiv igv_{ij}(\unx)(t_a)_{jk}
{\d\over\d v_{ik}(\unx)}=ig\tr\[v(\unx)t_a{\d\over\d v^T(\unx)}\],
\labels{opd}\ee
and the functional derivative in $v$ is defined so that 
\be
{\d v_{pq}(\un y)\over\d v_{ij}(\unx)}=\d_{pi}\d_{qj}\d(\unx-\un y).\ee

The inner product between two functionals of $v$ is defined to be
\be
\bk{\ov f(v)|\ov g(v)}=\int \ov f^*(v)\ov g(v)\D_H[v].\labels{innerv01}\ee
It can be shown (see Appendix A) that $\widehat \D_a(\unx)$
is anti-hermitean with respect to this inner product.
In particular, 
\be
\widehat \D_0(\unx)=
igv_{ij}({\unx})[\d/\d v_{ij}({\unx})]/\sqrt{2N_c}\nn\ee
is anti-hermitean, so the operator $\ov \D$ defined below
eq.~\eq{innergp} is hermitean, as previously claimed.

Instead of $\widehat\D_a({\unx})$ in \eq{opd}, it is  more 
convenient to deal with the matrix operators $\D(\unx)$ and
$\D'(\unx)$, whose $(mn)$ matrix
elements are defined to be
\be
\D_{mn}(\unx)&\equiv& {2\over ig}(t_a)_{mn}\widehat D_a({\unx})=
v_{in}({\unx}){\d\over\d v_{im}(\unx)},\nn\\
\D'_{mn}(\unx)&\equiv&v_{mi}({\unx}){\d\over\d v_{ni}({\unx})}.
\labels{dmat}\ee
It can be checked that these two are
related by
\be
v(\unx)\D(\unx)\vd_\unx(\unx)=\D'(\unx),\labels{ddp}\ee
where the subscript $\unx$ in $\vd_\unx(\unx)$ indicates
that this $\vd(\unx)$ should not be differentiated by
the $\d/\d v(\unx)$ in $\D(\unx)$. In other words, in component
forms, \eq{ddp} reads $v_{im}\vd_{nj}\D_{mn}=\D'_{ij}$.

When operated on $v_{pq}$ and $\vd_{pq}=v^{-1}_{pq}$, they yield
\be
\D_{mn}(\unx)v_{pq}(\un y)&=&+\d(\unx-\un y)\d_{mq}v_{pn}(\un y),\nn\\
\D_{mn}(\unx)\vd_{pq}(\un y)&=&-\d(\unx-\un y)\d_{np}\vd_{mq}(\un y),\nn\\
\D'_{mn}(\unx)v_{pq}(\un y)&=&+\d(\unx-\un y)\d_{np}v_{mq}(\un y),\nn\\
\D'_{mn}(\unx)\vd_{pq}(\un y)&=&-\d(\unx-\un y)\d_{mq}\vd_{pn}(\uny).
\labels{donv}\ee
These formulas give rise to the following formulas which are very
useful in practical calculations. When $\D$ or $\D'$ operates on a $v$
or $\vd$ in a trace, we have
\be
\D(\unx)\tr\[Av(\uny)\]&=&+\d(\unx-\uny)\[Av(\uny)\],\nn\\
\D(\unx)\tr\[\vd(\uny)A\]&=&-\d(\unx-\uny)\[\vd(\uny)A\],\nn\\
\D'(\unx)\tr\[v(\uny)A\]&=&+\d(\unx-\uny)\[v(\uny)A\],\nn\\
\D'(\unx)\tr\[A\vd(\uny)\]&=&-\d(\unx-\uny)\[A\vd(\uny)\].\labels{form1}
\ee
In other words, when the trace is written in a certain order, the operators
$D(\unx)$ and $D'(\unx)$ simply remove the trace, and append to it
the factor $\pm\d(\unx-\uny)$. When $\D$ or $\D'$ operates on a $v$
or $\vd$ in the same trace, we get
\be
\tr\[A\D(\unx)Bv(\uny)\]&=&+\d(\unx-\uny)\tr\[A\]\tr\[Bv(\uny)\],\nn\\
\tr\[A\D(\unx)B\vd(\uny)\]&=&-\d(\unx-\uny)\tr\[B\]\tr\[\vd(\uny)A\],\nn\\
\tr\[A\D'(\unx)Bv(\uny)\]&=&+\d(\unx-\uny)\tr\[B\]\tr\[v(\uny)A\],\nn\\
\tr\[A\D'(\unx)B\vd(\uny)\]&=&-\d(\unx-\uny)\tr\[A\]\tr\[B\vd(\uny)\].
\labels{form2}\ee
In other words, the single trace is broken up into a product of two 
traces. The matrices $A,B$ in these equations are constant matrices.

\subsection{\bm $\ov W_\t(v)$\ubm\ and its JIMWLK equation}
Using \eq{map}, the distribution function $W_\t(\a)$ can be
folded into the distribution function $\ov W_\t(v)$
of Wilson lines. Since
$\d/\d\a_\t^a({\unx})$ is transformed into 
$\widehat D_a({\unx})$ of \eq{opd},
the JIMWLK equation \eq{jimwlk1} for $W_\t(\a)$ is now changed 
into a JIMWLK equation for $\ov W_\t(v)$:
\be
{\p \ov W_\t(v)\over\p\t}&=&-\ov H\ \ov W_\t(v),\labels{jimwlkv1}\ee
where
\be
\ov H&=&{1\over\pi}\int d^2\unz d^2\unx d^2\uny 
K(\unx\un y|\un z)\ov\O(\unx\un y|\un z),
\nn\\
\ov\O(\unx\un y|\un z)&=&
\widehat D_a({\unx})\(V^\dagger(\unx)-V^\dagger(\un z)
\)_{ac}\(V(\un y)-V(\un z)\)_{cb}\widehat D_b({\un y}).\nn\\
\labels{jimwlkv2}\ee
Like $H$, $\ \ov H$ is also hermitean and positive semi-definite.

Using \eq{dmat} and \eq{vv2},
the operator $\ov\O(\unx\un y|\un z)$ can be written in
a form more convenient for practical calculations. 
Remember for this purpose the remark after \eq{gluewil4}, that 
the factor $[V^\dagger(\unx)-V^\dagger(z)]_{ac}$ in \eq{jimwlk2}
can be written to the left of the differential operator $\d/\d\a_\t^a(\unx)$.
In terms of \eq{jimwlkv2}, this means to the left of $\widehat\D_a(\unx)$.
In what follows we shall use $\ov\O(\unx\un y|\un z)$ of \eq{jimwlkv2}
in this form.

Using \eq{dmat}, \eq{ddp} and \eq{vv2}, and this remark above, we have
\be
V_{cb}(\un z)\widehat\D_b(\uny)
&=&ig\tr\(\vd(\un z)t_cv(\un z)\D(\uny)\),\nn\\
\widehat\D_a(\un x)V^\dagger(\un z)
&=&ig\tr\(\vd(\un z)t_cv(\un z)\D(\un x)\),\nn\\
V_{cb}(\un y)\widehat\D_b(\uny)
&=&ig\tr\(\vd(\un y)t_cv(\un y)\D(\uny)\)=ig\tr\(t_c\D'(\uny)\),\nn\\
\widehat\D_a(\un x)V^\dagger(\un x)
&=&ig\tr\(\vd(\un x)t_cv(\un x)\D(\un x)\)=ig\tr\(t_c\D'(\unx)\).
\labels{newop}\ee
From these relations, and the identity $\tr(t_cA)\tr(t_cB)=\h\tr(AB)$,
we get
\be
\ov\O(\unx\un y|\un z)&=&\ov\O_{xy}+\ov\O_{xz}+\ov\O_{zy}+\ov\O_{zz},\nn\\
\ov\O_{xy}&=&-\h g^2\tr\[\D'(\unx)\D'(\uny)\],\nn\\
\ov\O_{xz}&=&+\h g^2\tr\[\D'(\unx)v(\unz)\D(\uny)\vd_\uny(\unz)\],\nn\\
\ov\O_{yz}&=&+\h g^2\tr\[v(\unz)\D(\unx)\vd_\unx(\unz)\D'(\uny)\],\nn\\
\ov\O_{zz}&=&-\h g^2\tr\[v(\unz)\D(\unx)\vd_\unx(\unz)v(\unz)\D(\uny)
\vd_\uny(\unz)
\]\nn\\
&=&-\h g^2\tr\[\D(\unx)\D(\uny)\].\labels{oij}\ee

Assuming $\ov W_\t(v)$ to be normalized, $\int\D_H[v]\ov W_\t(v)=1,$
the average of any functional $B(v,\vd)$ is equal to
\be
\bk{B}_{_\t}=\int\D_H[v]B(v,\vd)\ov W_\t(v).\labels{avbar}\ee
If $B$ and $\ov W_\t$ are both monomial functionals of 
$v$ and $\vd$, this functional integral factorizes into a product
of integrals on the group $U(n)$, each of which can be computed
using \eq{ortho} and \eq{group}. In particular, the functional
integral is non-zero only when the number of $v$'s and $\vd$'s
in $B\ov W_\t$ are the same for every transverse position $\unx$.
This is the functional form of the 
{\it matching rule} previously considered.

\section{Mueller's Derivation of the JIMWLK Equation}
In a recent paper \cite{AM5}, Mueller proposed a simple derivation of the JIMWLK
equation in the following way. He showed that if $W_\t(\a)$ is equal 
to the dipole
functional $\vd(\uns)_{ij}v(\unt)_{kl}$, then the Feynman diagrams
for $\p W_\t(\a)/\p\t$ can be written in the form $-HW_\t(\a)$, with
$H$ given by \eq{jimwlk2}. He then stated that the same is true if $W_\t(\a)$
is equal to any multipole functional
$\vd(\uns_1)_{ i_ij_1}\vd(\uns_2)_{i_2j_2}\cdots
v(\unt_1)_{k_1l_1}v(\unt_2)_{k_2l_2}\cdots$, and hence it is likely that
the JIMWLK equation for $W_\t(\a)$ is also valid.

In this section we shall supply the missing steps
of this proof. This consists of filling in the detailed arguments
for the multiple functionals, and then showing that they lead to
the JIMWLK equation for $\ov W_\t(v)$. We do not know whether
the JIMWLK equation for $W_\t(\a)$ follows or not. However, for all
our applications, a JIMWLK equation for $\ov W_\t(v)$ is sufficient, so
it really does not matter whether the equation for $W_\t(\a)$
can be proven this way or not.

Instead of starting from the Feynman diagrams
to derive the evolution equation, as is done in Ref.~\cite{AM5}, we find
it easier to do everything in reverse. That is,
we start from the JIMWLK equation and show that they lead to
the correct set of Feynman graphs. This inverse approach makes it more
manageable to deal with the complicated multipole functionals. 
Actually, simplification already occurs at the
dipole level: a necessary cancelation 
in the original derivation is avoided altogether in this way.

The Wilson line $\vd(\uns)$ for a quark is drawn in Fig.~1 as a left-pointing
arrow, and the Wilson line $v(\unt)$ for an anti-quark is drawn as a 
right-pointing arrow. Time $x^-$ is drawn to increase from right to left; 
multiplication of color matrices from left to right should be carried out
against the arrow of the fermion lines.

The short vertical bars in the 
middle of the lines (labelled IP)
indicate the lightcone longitudinal position $x^-=0$ 
where interaction between the multipole and the pancake
nucleus takes place. Since $\a(\v x)$ is concentrated around
$x^-=0$, we may regard $v$ and $\vd$ to be located at the IP.

The operation of $\widehat D_a(\unx)$ or $\d/\d\a_\t^a(\unx)$ on the Wilson lines
is given in \eq{gluewil4}. This operation
can be represented graphically by putting a vertex to the left (the larger
$x^-$ side) of IP, both for the quark Wilson line $\vd$
and the anti-quark Wilson line $v$. The vertex for a quark is $igt_a$, and
the vertex for an anti-quark is $-igt_a$.

Using the remark following eq.~\eq{gluewil4},
the operator $\O(\unx\uny|\unz)$ in \eq{jimwlk2} can be written in the form
\be
\O&=&\[V^\dagger(\unx)-V^\dagger(\un z)\]_{ac}{\d\over\d \a_\t^a(\un x)}\.
\[V(\un y)-V(\un z)\]_{cb}{\d\over\d\a_\t^b(\un y)}\nn\\
&\equiv& \O_{\unx\uny}+\O_{\unx\unz}+
\O_{\unz\uny}+\O_{\unz\unz}.\labels{newo}\ee

Using \eq{vv1},
and the unitarity relation $v\vd=\vd v=1$, we see that the effect of
$\[V(\un y)\]_{cb}(\d/\d\a_\t^b(\un y))$, 
operating on a Wilson line, is to
place a vertex to the right (the small $x^-$ side) of the IP on the Wilson
line. Similarly, the effect of
$\[V^\dagger(\un x)\]_{ac}(\d/\d\a_\t^a(\un x))$, 
operating on a Wilson line, 
is also to
place a vertex to the right (the small $x^-$ side) of the IP on the Wilson
line. 

We are now ready to see what happens when
\be
H&=&\int d^2\unz d^2\unx d^2\uny
K(\unx\un y|\un z)\(\O_{\unx\uny}+\O_{\unx\unz}+
\O_{\unz\uny}+\O_{\unz\unz}\)\nn\\
&\equiv& H_{\unx\uny}+H_{\unx\unz}+H_{\unz\uny}+H_{\unz\unz}\labels{hxyz}\ee
operates on a multiple functional, {\it i.e.,}
a collection of $p$ quark and $q$ anti-quark Wilson lines. 

$H_{\unx\uny}$ puts a vertex
$\un y$ to the right of the IP on a Wilson line, 
and  vertex $\un x$ 
also to the right of the IP of the same or a different Wilson line.
 These two vertices are then connected by the `gluon propagator' 
$\int d^2\unz K(\unx\un y|\unz)\d_{ab}$,
where $a,b$ are the color indices at the two vertices. 
The gluon propagator is shown in Fig.~1(a) with a dashed line.
This operation is to be applied to every pair
of Wilson lines, including the possibility of applying to the same line
twice. 

Similarly, since $V^\dagger(\unz)V(\unz)=1$, the effect of 
$H_{\unz\unz}$ is to put
a vertex $\un y$ to the left of the IP of
a Wilson line, and another
vertex $\un x$ to the left of the IP of the same or another Wilson line. 
The two vertices are then linked by the same `gluon propagator'.
 This is shown in Fig.~1(b).

In both of these cases, the two vertices are both
 to the same side of the IP.
This is not the case with the other two terms.

$H_{\unx\unz}$ puts a vertex $\un y$ with color $b$
to the left of the IP, and a vertex $\un x$ with color $a$
to the right of the IP. These two vertices  
are linked by a `gluon propagator' 
$\int d^2\unz\[V(z)\]_{ab}K(\unx\uny|\un z)$. This is shown
in Fig.~1(c).

There is potentially another contribution to $H_{\unx\unz}$
when $V^\dagger_{ac}(x)\d/\d\a_a(x)$ operates on $V_{cb}(z)$.
However, this term is proportional to $\d(\unx-\unz)V^\dagger_{ac}(x)
V_{cd}(z)(T_a)_{db}$, which is proportional to
$(T_a)_{ab}=0$, so that term is actually absent.

Similarly, $H_{\unz\uny}$ puts a vertex $\un y$ 
of color $b$ to the right
 of the IP, and a vertex $\un x$ with color $a$
to the left of the IP. These two vertices
 are linked by the `gluon propagator'
 $\int d^2\unz\[V^\dagger(z)\]_{ab}K(\un x\un y|\un z)$. 
This is shown in Fig.~1(d).

Appropriate signs and Dirac $\d$-functions on the transverse
coordinates must also be inserted.

One might be bothered that the gluon propagators in the four
diagrams appear to be different. Fortunately this is only superficial. To 
see why they are actually the same, first note that $T_a$ is imaginary 
so $V$ is real.
Hence the gluon Wilson line $[V^\dagger(\unz)]_{ab}$ 
in the propagator
in Fig.~1(d) can be written as $[V(\unz)]_{ba}$. If we compare this
with that of Fig.~1(c), we see that these two are the same, both
equal to $[V(\unz)]_{a_1a_2}$, where $a_1$ is the color index before the
interaction point IP, and $a_2$ is the color index after.
Since $\a^a(x^-,\unz)$ is concentrated near $x^-=0$, we may write,
in both cases, the gluon Wilson line to be
\be
[V(\unz)]_{a_1a_2}=\widetilde P\exp\[-ig\int_{x_1^-}^{x_2^-}dx^-
\a^a(x^-,\unz)T_a\],\labels{ngwl}\ee
where $(a_1,x_1^-)$ is the interaction vertex to the right of
IP, and $(a_2,x_2^-)$ is the one to the left of IP.
For the first two diagrams, we may replace $\d_{ab}$ by the same
expression \eq{ngwl}, because in that case both $x_1^-$ and
$x_2^-$ are to the same side of IP, hence
$\a^a(x^-,\unz)=0$ throughout the integration interval, so
$[V(\unz)]_{ab}=\d_{ab}$.

These four types of Feynman graphs are precisely those
needed for the development $\p W_\t(\a)/\p\t$
\cite{AM5}. Therefore the JIMWLK
equation is satisfied whenever
 $W_\t(\a)$ is given by a multipole function, namely, a monomial
functional of $v$ and $\vd$.
Since polynomial functionals on the group manifold
form an orthonormal complete set, the JIMWLK equation
\eq{jimwlkv1} must be true in general. This completes Mueller's
proof for $\ov W_\t(v)$. However,
since $W_\t(\a)$ depends on many more variables
$\a_a(x^-,\unx)$
than $v(\unx)$, it does not necessarily follow from this argument that
\eq{jimwlk1} for $W_\t(\a)$ is true.

\section{Infrared Divergence and the Generalized BK Equation}
The kernel $K(\unx\uny|\unz)$ of the JIMWLK equation, given
in \eq{jimwlk3}, goes like $ 1/\unz^2$ for large $|\unz|$.
This causes a log divergence in $ H W_\t(\a)$ of \eq{jimwlk2}, and
$\ov H\ \ov W_\t(v)$ of \eq{jimwlkv2}. On the other hand,
the dipole amplitude \eq{satdipole} satisfies the Balitsky
equation \eq{balitsky}, whose kernel goes like $1/(\unz^2)^2$
for large $|\unz|$, so infrared divergence is absent in that case.
In this section, we shall use the Feynman diagrams derived 
in the last section to show
that the multipole traces defined below
are equally free of infrared divergence.

{\it Multipole traces} are defined by
\be
m(\uns_1\unt_1\cdots\uns_k\unt_k)\equiv{1\over N_c}\tr\[
\vd(\uns_1)v(\unt_1)\cdots \vd(\uns_k)v(\unt_k)\].\labels{ma}\ee
We will call the functional average of a multiple trace,
$\bk{m_k(\uns_1\unt_1\cdots\uns_k\unt_k)}$, a {\it multiple amplitude}.
When $k=1$, this reduces to the dipole amplitude \eq{dipole}.

The diagram for a multiple trace is shown in Fig.~2. 
As far as color-matrix multiplication is concerned, these $2k$
lines should be  considered to be joined together at
their ends to form a single
big loop with the arrows all pointing the
same way. The joints are indicated by dotted lines in the figure.
 Note that if the amplitude did not have the form displayed in
\eq{ma}, with $\vd$ and $v$ occuring alternately
inside a trace, such a big loop cannot be formed and the argument
below will not be valid.

Consider the two gluon propagators shown in Fig.~2. The color
structures are identical, but the two terms differ by a sign
because the vertex B is on a quark line in one diagram and on
an anti-quark line in another diagram. Their gluon propagators
may therefore be combined into
$K(\unx\unx|\unz)-K(\unx\uny|\unz)$. For large $|\unz|$, this is
proportional to $\unz\.(\unx-\uny)/(\unz^2)^2=O(1/|\unz|^3)$,
hence the infrared divergence is absent. As a matter of fact,
we can also combine diagrams with vertex A similarly shifted.
The four $K$ thus combined actually dies down like $1/(\unz^2)^2$
for large $|\unz|$.

It is clear that all the diagrams for the multipole function
can be paired up in a way similar to Fig.~2, thus eliminating
all infrared divergences. Moreover, since the combination
is obtained by combining two diagrams with vertex
B in different positions, but the same vertex A, 
finiteness persists for products of multipole traces.

In the rest of this section, we shall derive a generalization of the 
Balitsky-Kovchegov (BK) equation to multipole amplitudes. 
We will see that the kernel
of the equation actually behaves like $1/(\unz^2)^2$ for large $|\unz|$,
just like the kernel for the Balitsky equation.

Since $\ov H$ is hermitean, the JIMWLK equation satisfied by $\ov W_\t(v)$
is also satisfied by the multipole trace \eq{ma}. We may now use
\eq{form1}, \eq{form2}, \eq{jimwlkv2} and \eq{oij} to simplify
$\ov Hm(\uns_1\unt_1\cdots \uns_k\unt_k)$. The resulting equation is
\be
{\p m(\uns_1\unt_1\cdots \uns_k\unt_k)\over\p\t}=
-\[H_{\unx\uny}+H_{\unz\unz}+H_{\unx\unz}+H_{\uny\unz}\]
m(\uns_1\unt_1\cdots \uns_k\unt_k),\labels{gb1}\ee
where
\be
H_{\unx\uny}m(\uns_1\cdots\unt_k)&=&+\h g^2N_c
\sum_{i,j=1}^k\int d^2\unz I_{ij}(\unz)
m_{ij}^{aa}(\uns_1\cdots \unt_k)\nn\\ 
H_{\unz\unz}m(\uns_1\cdots\unt_k)&=&+\h g^2N_c
\sum_{i,j=1}^k\int d^2\unz I_{ij}(\unz)
m_{ij}^{bb}(\uns_1\cdots \unt_k)\nn\\
H_{\unx\unz}m(\uns_1\cdots\unt_k)&=&-\h g^2N_c
\sum_{i,j=1}^k\int d^2\unz I_{ij}(\unz)
m_{ij}^{ba}(\uns_1\cdots \unt_k)\nn\\ 
H_{\unz\uny}m(\uns_1\cdots\unt_k)&=&-\h g^2N_c
\sum_{i,j=1}^k\int d^2\unz I_{ij}(\unz)
m_{ij}^{ab}(\uns_1\cdots \unt_k).\labels{gb2}\ee
The kernel in these equations is
\be
I_{ij}(\unz)=K(\uns_i\uns_j|\unz)-K(\uns_i\unt_j|\unz)-K(\uns_j\unt_i|\unz)
+K(\unt_i\unt_j|\unz),\labels{gb3}\ee
which is $O(1/(\unz^2)^2)$ for large $|\unz|$. It is also symmetric in $i$
and $j$: $I_{ij}(\unz)=I_{ji}(\unz)$.
The meaning of the multipole traces in \eq{gb1} will now be explained.

The argument $(\uns_1\cdots\unt_k)$ of the 
multiple trace in \eq{ma} is circular, because the
trace is. We can consider the argument $\uns_1$ to be behind the
argument $\unt_k$, forming a circle. 
For example, 
\be
m(\uns_1\unt_1\cdots\uns_k\unt_k)=m(\unt_1\uns_2\cdots\unt_k\uns_1)
\equiv{1\over N_c}\tr\[v(\unt_1)\vd(\uns_2)\cdots v(\unt_k)\vd(\uns_1)\].\ee
In these formulas,
$v$ is always associated with a $\unt_i$ and $\vd$ is always
associated with a $\uns_i$. 

The quantity $m_{ij}^{aa}(\uns_1\cdots\unt_k)$ appearing in the
$H_{\unx\uny}$ term in \eq{gb2} is defined as follows. Put a vertical bar
{\it after} $\uns_i$, and another vertical bar {\it after}
$\uns_j$ in the {\it circular} argument $(\uns_1\cdots\unt_k)$.  
This pair of vertical bars separates the circular argument
into two circular arguments.
 $m_{ij}^{ab}(\uns_1\cdots\unt_k)$
is defined to be the product of two multiple traces
with these two circular argument. For example,
\be
m_{13}^{aa}(\uns_1\unt_1\uns_2\unt_2\uns_3\unt_3)\equiv
m(\uns_1|\unt_1\uns_2\unt_2\uns_3|\unt_3)\equiv
m(\unt_3\uns_1)m(\unt_1\uns_2\unt_2\uns_3).\ee

The superscript $a$ stands for `after'. The superscript $b$ to be
found in the other three terms of \eq{gb2} stands for `before'.
Each superscript together with its corresponding subscript pair
up to tell us where the vertical bar is put. The pair $(ai)$ means
to put a vertical bar after $\uns_i$. The pair $(bj)$ tells us to
put a vertical bar before $\uns_j$. With this understanding the
quantity $m^{bb}_{ij}(\uns_1\cdots\unt_k)$ can be defined similarly.
For example,
\be
m_{13}^{bb}(\uns_1\unt_1\uns_2\unt_2\uns_3\unt_3)\equiv
m(|\uns_1\unt_1\uns_2\unt_2|\uns_3\unt_3)\equiv
m(\uns_1\unt_1\uns_2\unt_2)m(\uns_3\unt_3).\ee

If we apply these recipes literally to the last two equations in
\eq{gb2} we will end up with something non-sensical, in that
the numbers of $v$'s and $\vd$'s within each trace is not identical,
thus neither is a multipole trace. What we should do in these two
cases is to insert a $v(\unz)$ into one trace and a $\vd(\unz)$
into another so that both become multiple functionals. For example,
\be
m_{13}^{ab}(\uns_1\unt_1\uns_2\unt_2\uns_3\unt_3)\equiv
m(\uns_1|\unt_1\uns_2\unt_2|\uns_3\unt_3)\equiv
m(\unt_1\uns_2\unt_2\unz)m(\unz\uns_3\unt_3\uns_1).\ee
The $\unz$ in the first factor is a $\vd(\unz)$, and that in the second
factor is a $v(\unz)$.

This completes the explanation of the symbols in \eq{gb2}, except for
one last remark. If the two vertical bars are side by side, then
the corresponding mulitple trace $m$ should be interpreted as
$\tr({\bf 1})/N_c=1$.

When we take expectation values on both sides of \eq{gb1}, we get
an equation for the multiple amplitudes. For large $N_c$, the
expectation of products of traces factorizes into products of 
expectations of a trace. In this form \eq{gb1} and \eq{gb2} 
generalize the Balitsky-Kovchegov (BK)
 equation to higher multiple amplitudes. We should interpret
$m$ in these equations as the average $\bk{m}$, and products of
two $m$'s as the product of the averages.

In the special case of a dipole, $k=1$, hence $i=j=1$ in
\eq{gb2}. In that case, 
\be
m^{aa}_{11}&=&m(\uns_1||\unt_1)=m(\uns_1\unt_1),\nn\\
m^{bb}_{11}&=&m(||\uns_1\unt_1)=m(\uns_1\unt_1),\nn\\
m^{ba}_{11}&=&m^{ab}_{11}=m(|\uns_1|\unt_1)=m(\uns_1\unz)m(\unz\unt_1),\nn\\
I_{11}(z)&=&{1\over 4\pi^3}{(\uns_1-\unt_1)^2\over(\uns_1-\unz)^2(\unt_1-\unz)^2}.\ee
Equation \eq{gb1} coincides with the BK equation
\eq{kovchegov}, as it should.

A very interesting fact emerges from these generalized BK
equations for multipole amplitudes. If the multipole amplitudes
for $k=1,2,\cdots,\ell-1$ are known, then the evolution equation 
determining the $\ell$th multipole is a {\it linear} inhomogeneous
equation. Thus the only non-linear equation one has to solve is the 
original BK equation for the dipole
amplitude. It is well known that the non-linearity of this equation
leads to saturation. The linearity of the higher multipole amplitudes
therefore means saturation of the dipole amplitude automatically
drives saturation of all higher multipole amplitudes.

If we require the solution of the JIMWLK equation \eq{jimwlkv1}
to be free of infrared singularitities, then presumably it will be
made out of the multiple traces \eq{ma} and their products. In that
case the difficult functional differential JIMWLK equation can be
replaced by the more manageable set of generalized BK equations
\eq{gb1} and \eq{gb2}.

\section{Solution of the JIMWLK Equation}
Suppose we decompose $\ov W_\t(v)$ into a linear combination of eigenfunctions
$\ov\phi_\l(v)$ of $\ov H$. The operator
$\ov H$ is hermitean
and positive semi-definite, hence the eigenvalues
 $\l$ are real and non-negative.
From \eq{jimwlkv1}, the $\t$ dependence of $\ov W_\t(v)$ is thus
given by a linear combination of $\exp(-\l\t)$. In the asymptotic
limit  $\t\to\infty$, the lowest eigenvalue of $\ov H$ dominates.

The lowest eigenvalue is $\l=0$,
and its normalized eigenfunction is $\ov \f_0(v)=1$.
This eigenfunction is normalized because $\int\D_H[v]=\prod_\unx\int
d_H[v(\unx)]=1$.
 
In contrast, $\f_0(\a)=1$ is also an eigenfunction of $H$ with
$\l=0$, but the integral of this eigenfunction is divergent because
the linear space $\A$ is non-compact.

In the next subsection, we will discuss what happens in the asymtotic
 limit $\t\to\infty$
 when $\l=0$ dominates. In the subsequent subsection,
we will look at $\ov W_\t(v)$ and its applications for a smaller $\t$.

\subsection{Asymptotic limit}
When $\t=\infty$, only $\ov\phi_0(v)$ contributes, so we can set
$\ov W_\infty(v)=\ov\f_0(v)=1$. 
Averages are then given by the integral $\bk{B(v,\vd)}=\int\D_H[v]B(v,\vd)$.
From the matching rule,  this integral  is non-zero only when
the number of $v$'s in $B$ exactly matches the number of $\vd$'s 
at every transverse position $\unx$.

For example, the dipole amplitude \eq{dipole}
is 
\be
s_\t(\unx,\un y)={1\over N_c}\bk{\tr\[\vd(\unx)v(\un y)\]}
=\d_{\unx,\un y}\labels{satdipole}\ee
because $\vd_\unx v_\unx=1$ and $\tr(1)=N_c$. 
This is consistent with \eq{normdip}, and also \eq{larges2}
at $\t=\infty$. In other words, whenever the dipole has
a finite size, the dense gluon will have such a strong
absorption that the dipole amplitude always become zero.

A similar statement can be made about multipole amplitudes.

Let us now compute the gluon spectrum at $\t=\infty$ by using the formula
\cite{EI1}
\be
{d(x_{_F}G)\over d^2\un k}&=&{1\over 4\pi^3}\bk{F^{+i}_a(\v k)
F^{+i}_a(-\v k)}={1\over 4\pi^3N_c}\bk{\Tr\[F^{+i}(\v k)F^{+i}
(-\v k)\]},
\labels{glue1}\ee
where $\v k=(k^+,\un k)$, and 
\be
F^{+i}_a(\v k)&=&\int d^3\unx \exp(i\unk\.\unx)F^{+i}(\v x)
\labels{glue2}\ee
is the color electric field in the lightcone gauge. In the approximation
$F^{+i}(\v x)\simeq (i/g)\d(x^-)V(\unx)\p^iV^\dagger(\unx)$ 
which is supposed to be valid for a 
Lorentz-contracted pancake nucleon, the formula becomes
\be
{d(x_{_F}G)\over d^2\un k}&=&-{1\over 4\pi^3g^2N_c}\int d^2\unx d^2\uny
\exp[i\un k\.
(\unx-\un y)]\bk{C(\unx,\uny)},\nn\\
C(\unx,\uny)&=&\Tr\[V(\unx)V^\dagger_i(\unx)V(\un y)V^\dagger_i(\un y)\],
\labels{glue3}\ee
where the subscript $i$ represents a differentiation, namely,
 $V^\dagger_i\equiv\p_iV^\dagger$. 

To compute the average 
$ \bk{C(\unx,\uny)}=\int \D_H[v]B(\unx,\uny)$, 
we need to use \eq{vv2} to convert the gluon 
Wilson lines $V$ and $V^\dagger$
to the quark and anti-quark Wilson lines $\vd$ and $v$. The result is
\be
C(\unx,\uny)&=&2C_F\tr\[v(\uny)\vd_i(\uny)v(\unx)\vd_i(\unx)+v_i(\uny)\vd(\uny)
v_i(\unx)\vd(\unx)\]\nn\\
&+&\tr\[v_i(\uny)\vd(\uny)\]\tr\[v(\unx)\vd_i(\unx)\]\nn\\
&+&\tr\[v(\uny)\vd_i(\uny)\]\tr\[v_i(\unx)\vd(\unx)\].\labels{bsmallv}\ee
The functional integral $\bk{C}=\int \D_H[v]C$ is computed in Appendix B.
The result is
\be
\bk{C(\unx,\uny)}&=&-2{N_c^2-1\over a^2}\d_{\unx,\uny}.\labels{avb}\ee
Changing the integral in \eq{glue3} into a finite sum, and letting
$\un r=\unx-\uny$, we get
\be
{d(x_{_F}G)\over d^2\unk}&=&{\pi R_A^2\over 4\pi^3g^2N_c}
\sum_{\un r}a^2\exp(i\unk\.\un r){2(N_c^2-1)\over a^2}\d_{\un r,0}\nn\\
&=&{2\pi R_A^2(N_c^2-1)\over 16\pi^4\a_sN_c}.\labels{gluespec}\ee
The unintegrated spectrum $d(x_{_F}G)/d\unk^2$ is therefore absolutely flat,
up to the saturation momentum $Q_s^2$ which is in this case infinite.
That is not unexpected at $\t=\infty$.

The integrated cross section is then
\be
x_{_F}G=Q_s^2{\pi R_A^2(N_c^2-1)\over 8\pi^3\a_sN_c}.\labels{total}\ee

The spectrum in \eq{gluespec} differs from the mean field prediction
\eq{meanfield}, in that \eq{gluespec} is flat and \eq{meanfield}
has a logarithmic dependence on $\unk^2$. The integrated density
$x_{_F}G$ is however quite similar to the estimate given
in \eq{qs}; the only difference is a factor $c=8\pi^3$. 

\subsection{Below the asymptotic limit}
It is much more difficult to solve the JIMWLK equation 
for finite $\t$, because we know nothing about the other 
eigenfunctions and eigenvalues
of $\ov H$. In this section,
we will discuss an approximate correction
to $\ov W_\t(v)$ below the asymptotic limit, in a region
where the Wilson lines are far apart. 

To avoid the infrared divergence, the distribution functional $\ov W_\t(v)$
will be assumed to depend on $v$ and $\vd$ only through the multipole
traces \eq{ma}, or products of them. 
From results of the last subsection, and discussions in Sec.~2, we know
that for large $\t$, the average of multipole traces
 ({\it i.e.,} multipole amplitudes) are small 
if the Wilson lines in the multipoles are far apart.
In fact, the higher the order of the multipole is, the smaller the
amplitude is expected to be. Therefore it is reasonable to include
only quadratic dependences of $v$ and $\vd$
in a first correction to the asymptotic limit, at least
in the region when
the Wilson lines are far apart.
We will therefore assume 
\be
\ov W_\t(v)=1+\sum_{\uns,\unt}\tr\[\vd(\uns)v(\unt)\]b_\t
(\unt,\uns).\labels{b}\ee

The contribution  $\tr[\vd(\uns)v(\uns)]b_\t(\uns,\uns)=N_cb_\t(\uns,\uns)$ 
may be 
absorbed into the $v$-independent term. We may therefore assume
$b_\t(\uns,\uns)=0$ without any loss of generality.
In that case, using the matching rule and unitarity of $v$
to do the functional integral, we see that $\ov W_\t(v)$ is
still normalized:
\be
\int \ov\D_H[v]\ W_\t(v)=1.\ee

To compute the dipole amplitude \eq{dipole}, we need the following 
integration formula, 
which can be obtained
 from the matching rule, eq.~\eq{vv}, and the unitarity of $v$:
\be
\int \D_H[v]\ \tr\[\vd(\uns)v(\unt)\]\tr\[\vd(\unx)v(\uny)\]
=N_c^2\d_{\uns,\unt}\d_{\unx,\uny}+\d_{\uns,\uny}
\d_{\unt,\unx}-\d_{\uns,\uny,\unt,\unx}.\nn\\ \ee
The last Kronecker $\d$ is by definition
non-zero only when the four arguments
in its subscript are all equal.

We may now compute the dipole amplitude from \eq{dipole} to be
\be
s_\t(\unx,\uny)&=&
{1\over N_c}\int\D_H[v]\ \ov W_\t(v)\tr\[\vd(\unx)v(\uny)\]\nn\\
&=&\[\d_{\unx,\uny}+{1\over N_c}b_\t(\unx,\uny)\].
\ee
Since $b_\t(\unx,\unx)=0$, we get $s_\t(\unx,\unx)=1$, as it should be.
If we stay away from $\unx=\uny$, then $b_\t(\unx,\uny)/N_c$
is just the dipole amplitude $s_\t(\unx,\uny)$. As such it should
satisfy the BK equation. On the other hand, we
should be able to get the equation of $b_\t(\unx,\uny)$
directly from the JIMWLK equation \eq{jimwlkv1} by using 
\eq{b}, \eq{oij}, and \eq{donv}. The left hand side is proportional
to $\tr\[\vd(\uns)v(\unt)\]$, but the right hand side has two
terms, one is proportional to $\tr\[\vd(\uns)v(\unt)\]$, and the other
it proportional to $\tr\[\vd(\unz)v(\unt)\]\tr\[\vd(\uns)v(\unz)\]$.
If we drop this quartic term because it involves a higher order
multipole function which is expected to be small, then 
$b_\t(\unx,\uny)/N_c=s_\t(\unx,\uny)$
simply satisfies the BK equation \eq{kovchegov}
with the quadratic term of $s_\t$ dropped. This is justified
when the dipole amplitude is small, which is the case when
the two Wilson lines are far apart, as expected. The solution
is given by \eq{larges1}.

We may use \eq{b} to calculate higher-order amplitude. The result
is a sum of $b_\t/N_c=s_\t$, one for each dipole pairs inside
the multipole.

\section{Conclusion}
Density of soft gluons is determined by the distribution
$W_\t(\a)$ of the
 classical Yang-Mills potential $\a(\v x)$. Interaction of 
fast partons with such a background is given by their
Wilson lines. In this paper,
we introduced the distribution $\ov W_\t(v)$ of Wilson lines. It can be
obtained from $W_\t(\a)$, and it also
satisfies the JIMWLK equation. We completed
Mueller's derivation of the JIMWLK
equation, though for $\ov W_\t(v)$ and not for $W_\t(\a)$. 
We derived a generalized BK equation for multipole amplitudes.
We also used the normalizable property of $\ov W_\t(v)$ 
 to compute the properties of physical observables
at $x_{_F}=0$. We obtained in this way an unintegrated gluon spectrum
$d(x_{_F}G)/d\unk^2=
\pi R_A^2(N_c^2-1)/8\pi^3\a_sN_c$, independent of its transverse
momentum $\unk^2$ of the gluon. Correction to this asymptotic behavior
was briefly discussed.

\acknowledgements
This work of CSL is supported partly by the Natural Sciences and Engineering
Research Council of Canada, and the Fonds de recherche sur la nature et les
technologies of Qu\'ebec. The work of WZ is supported by the National
Natural Science Foundation of China, Grant Nos.~10075020 and 90103013.
He wants to express his gratitude to the hospitality of McGill University
where this work is carried out.

\appendix
\section{Mathematical details}

The Haar measure of $U(n)$ for odd $n$ is given by
\be
d_H[v]=c_4\ \tr\[v^{-1}dv\wedge v^{-1}dv\wedge\cdots\wedge v^{-1}dv\],
\labels{oddn}\ee
where $c_4$ is a normalization constant determined by $\int d_H[v]=1$,
and the argument of $\tr$ consists of the exterior product of $n^2$
factors of $v^{-1}dv$. From the cyclic nature of the trace, and the
antisymmetry character of the exterior product, it can be seen that
$d_H[v]$ defined this way is 0 when $n$ is even. Hence we must confine 
ourselves to an odd $n$ of we want to use the expression \eq{oddn}.

It is easy to check that \eq{haar} is satisfied for \eq{oddn}.

We will show in two ways that $\widehat\D_a(\unx)$
defined in \eq{opd} is anti-hermitean. First, define the inner product
of two functionals $\a$ by
\be
\bk{f|g}_\a&=&\int \D[\a]\D[\a']f^*(\a)g(\a')\d\(u-u'\)/\J(u),
\labels{innera}
\ee
where
$u$ is given by \eq{u} and $u'$ is given similarly in terms of $\a'$.
From \eq{dv} and \eq{dvda}, we see that
\be
\d(u-u')/\J(u)=\prod_{a,\unx}\d\(\a_\t(\unx)-\a'_\t(\unx)\).\ee
Using integration by parts and assuming the 
resulting surface term to be zero, it
is easy to see that $\d/\d\a_\t^a(\t,\uny)$ is anti-hermitean with respect
to this inner product. 

If $\ov f(v)$ and $\ov g(v)$ are
obtained respectively from $f(\a)$ and $g(\a)$ by \eq{map}, then 
the inner product between $\ov f(v)$ and $\ov g(v)$ defined by
\be
\bk{\ov f|\ov g}_v\equiv\int \D_H[v]\ov f^*(v)g(v)\labels{innerv}\ee
is equal to the inner product $\bk{f|g}_\a$ defined in \eq{innera}.
This is so because
\be
\bk{\ov f|\ov g}_v&=&\int \D_H[v]\D[\a]\D[\a']f^*(\a)\d(v-u)g(\a)
\d(v-u')/\J(u)\J(u')\nn\\
&=&\int \D[\a]\D[\a']\d(u-u')f^*(\a)g(\a')/J(u)=\bk{f|g}_\a.\ee

We know that the transform of $\d f(\a)/\d\a_\t^a(\uny)$ is 
$\widehat\D_a(\uny)$. Since $\bk{\ov f\ov g}_v=\bk{f|g}_\a$ and since
$\d/\d\a_\t^a(\uny)$ is anti-hermitean with respect to $\bk{f|g}_\a$,
it follows that $\widehat\D_a(\uny)$ must also be anti-hermitean
with respect to $\bk{\ov f|\ov g}$.

The second proof of the anti-hermiticity of $\widehat\D_a(\uny)$
makes use of the explicit formula \eq{oddn}. It proceeds as follows.

Using the explicit formula \eq{opd} for $\widehat\D_a(\uny)$, we must
show that $\ov\D_a(\uny)\equiv v_{ij}(\uny)(t_a)_{jk}\d/\d v_{ik}(\uny)$
is hermitean with respect to the inner product \eq{innerv}.

Let us first show that $\ov\D_a(\uny)$ is imaginary. 
Since $\vd=v^{-1}$, it follows that 
\be
{\d\over\d v^*_{ik}}=-v_{is}v_{rk}{\d\over\d v_{rs}}.\labels{inv}\ee
Using also the fact that $t_a$ is hermitean, then
\be
\ov\D_a^*&=&
v_{ij}^*(t_a^*)_{jk}{\d\over\d v^*_{ik}}
=-(vt_a)_{rs}{\d\over\d v_{rs}}
=-\ov\D_a.\labels{ima}\ee
Now we use integration by parts to show that $\ov\D_a$ is antisymmetric
with respect to the inner product $\bk{f|g}_v$. Remembering \eq{dv},
integration by parts changes $\ov\D_a$ into
\be
-{1\over\J(v)}{\d\over\d v_{ik}}(vt_a)_{ik}\J(v).\ee
This would indeed be equal to $-\ov\D_a$ if
\be
{\d\over\d v_{ik}}\[(vt_a)_{ik}\J(v)\]=0.\labels{rev}\ee
Since $\d\(d[v]\)/\d v_{ik}=d\[\d v/\d v_{ik}\]=0$,
\eq{rev} is true if
\be
{\d\[(vt_a)_{ik}\D_H[v]\]\over\d v_{ik}}=0,\ee
 which in turn is true if
\be
n\tr(t_a)d_H[v]+(vt_a)_{ik}{\p d_H[v]\over\p v_{ik}}=0.\labels{toshow}\ee
Using \eq{oddn}, we get
\be
(vt_a)_{ik}{\p d_H[v]\over\p v_{ik}}&=&-n^2\tr\[t_av^{-1}dv
\wedge\cdots\wedge v^{-1}dv\],\ee
namely, it is equal to $-n^2$ times a $t_a$ inserted into
the measure in \eq{oddn}. If $a=0$, then $t_0$ is proportional
to the identity matrix, so indeed \eq{toshow} is true.
For $a>0$, $\tr(t_a)=0$. Since
$v^{-1}dv$ is a linear combination of the $U(n)$ generators
$t_b$, we conclude that \eq{toshow} is valid if
\be
\e_{b_1b_2\cdots b_{n^2}}\tr\[t_at_{b_1}\cdots t_{b_{n^2}}\]=0.\ee  
This is indeed the case because $\tr(t_a)=0$ and 
$s\equiv \e_{b_1b_2\cdots b_{n^2}}t_at_{b_1}\cdots t_{b_{n^2}}$
is proportional to the identity matrix. This last statement can be
proven as follows.

Let $v_0\in U(n)$. Then $v_0^{-1}t_av_0=(V_0)_{ab}t_b$, where
$V_0$ is the adjoint representation of $v_0$. Hence
$v_0^{-1}sv_0=\det(V_0)s$. Since the adjoint generator $T_a$ is
has no diagonal matrix elements, it is traceless, so $\det(V_0)=1$.
Therefore $s$ commutes with every element $v_0$ of $U(n)$, so by
Schur's lemma, it must be proportional to the identity matrix.

We have thus shown that $\ov\D_a(\uny)$ is antisymmetric and imaginary.
Hence it is hermitean.

\section{Saturation calculation}
We want to compute $\bk{C(\unx\uny)}=\int\D_H[v]C(\unx,\uny)$ for
the function $C(\unx\uny)$ given in \eq{bsmallv}. This function
contains four terms. We will label them consecutively as
$C_1,C_2,C_3$, and $C_4$.

We interpret the derivative $\p_iv(\unx)=v_i(\unx)$ on the lattice
to be
\be
v_i(\unx)={1\over 2a}\[v(\unx+\un a_i)-v(\unx-\un a_i)\],\ee
where $\un a_i$ is a lattice vector along the $i$th direction.
$\vd_i(\unx)$ is defined similarly. The following computations make  use 
of \eq{ortho}, $C_F=N_c/2$ for $U(n)$, the matching rule
discussed at the end of Sec.~3, and unitarity of $v$. There are also
extra factors of 2 obtained by summing the two
$\un a_i$ directions. In this way we get
\be
\bk{C_1}&=&{2C_F\over 4a^2}\bk{\tr\bigg[v(\uny)\(\vd(\uny+\un a_i)-
\vd(\uny-\un a_i)\)\nn\\
&&\hskip1.5cm v(\unx)\(\vd(\unx+\un a_i)-\vd(\unx-\un a_i)\)\bigg]}
\nn\\
&=&-{2C_FN_c\over 4a^2}\[\d_{\unx,\uny-\un a_i}+\d_{\unx,\uny+\un a_i}\]
\simeq -{2C_FN_c\over a^2}\d_{\unx,\uny}\nn\\
&=&-{N_c^2\over a^2}\d_{\unx,\uny},\nn\\
\bk{C_2}&=&{2C_F\over 4a^2}\bk{\tr\bigg[\(v(\uny+\un a_i)-
v(\uny-\un a_i)\)\vd(\uny)\nn\\
&&\hskip1.6cm \(v(\unx+\un a_i)-v(\unx-\un a_i)\)\vd(\unx)\bigg]}
\nn\\
&=&\bk{C_1},\nn\\
\bk{C_3}&=&{1\over 4a^2}\tr\[\(v(\uny+\un a_i)-v(\uny-\un a_i)\)\vd(\uny)\]\nn\\
&&\hskip.7cm\tr\[v(\unx)\(\vd(\unx+\un a_i)-\vd(\unx-\un a_i)\)\]\nn\\
&=&{1\over a^2}\d_{\unx,\uny},\nn\\
\bk{C_4}&=&{1\over 4a^2}\tr\[v(\uny)\(\vd(\uny+\un a_i)-\vd(\uny-\un a_i)\)\]\nn\\
&&\hskip.7cm\tr\[\(v(\unx+\un a_i)-v(\unx-\un a_i)\vd(\unx)\)\]\nn\\
&=&{1\over a^2}\d_{\unx,\uny}.\nn\\
\ee
Thus
\be
\bk{C(\unx,\uny)}=\bk{C_1+C_2+C_3+C_4}=-2{N_c^2-1\over a^2}\d_{\unx,\uny}.\ee




\begin{figure}[h]

\vspace*{15cm}
\includegraphics{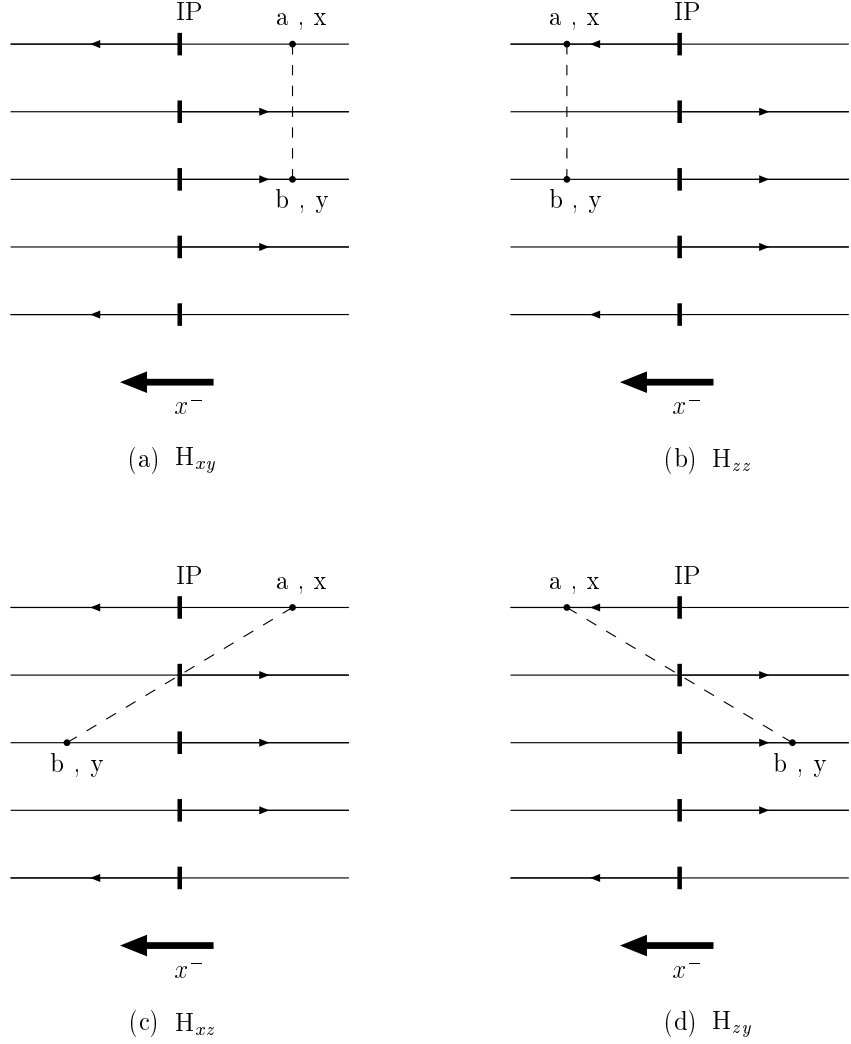}
\vspace*{2.5cm}
\caption[]{Diagrams representing the result of 
$H=H_{xy}+H_{zz}+H_{xz}+H_{zy}$, or 
$\ov H=\ov H_{xy}+\ov H_{zz}+\ov H_{xz}+\ov H_{zy}$,
operating on a multipole functional with 2 anti-quark Wilson lines $\vd$
(left-pointing arrows) and three quark Wilson lines
$v$  (right-pointing arrows). The dashed lines are the
`gluon propagators'.
See the text in Sec.~4. 
for further explanation.}
\end{figure}

\begin{figure}[h]

\vspace*{15cm}
\includegraphics{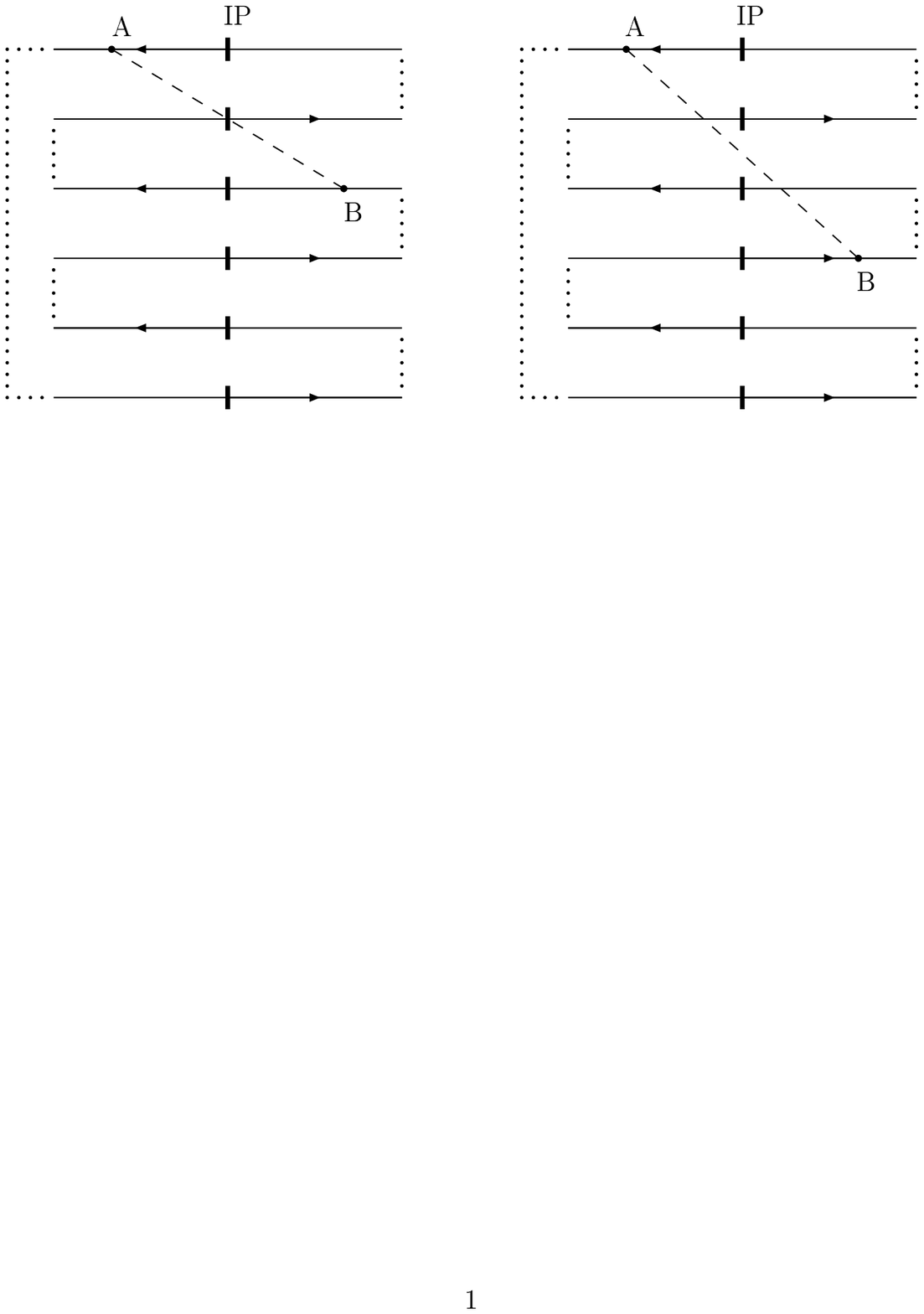}

\caption[]{A multiple trace with $k=3$. The two diagrams have
identical color structure, so their two `gluon propagators' 
can be combined
into an infrared finite expression.}
\label{fig2}
\end{figure}

\end{document}